\documentstyle[aps,twocolumn]{revtex}

\begin{document}
\title{Realization of All-or-nothing-type Kochen-Specker Experiment with Single
Photons}
\author{Yun-Feng Huang\thanks{%
Email address: hyf@mail.ustc.edu.cn}, Chuan-Feng Li, Yong-Sheng Zhang\thanks{%
Email address: yshzhang@ustc.edu.cn}, Jian-Wei Pan, and Guang-Can Guo\thanks{%
Email address: gcguo@ustc.edu.cn}}
\address{Key Laboratory of Quantum Information, University of Science and Technology\\
of China, Chinese Academy of Science, Hefei, Anhui, P. R. China, 230026}
\maketitle

\begin{abstract}
Using the spontaneous parametric down-conversion process in a type-I phase
matching BBO crystal as single photon source, we perform an
all-or-nothing-type Kochen-Specker experiment proposed by Simon {\it et al}.
[Phys. Rev. Lett. {\bf 85}, 1783 (2000)] to verify whether noncontextual
hidden variables or quantum mechanics is right. The results strongly agree
with quantum mechanics.

\end{abstract}

\baselineskip20ptThe problem of hidden variables in quantum mechanics has
been discussed for many years. In 1964, Bell derived his inequality\cite
{Bell} to show the contradiction between local hidden variables (LHV) and
quantum mechanics (QM). Bell inequality enabled experimental test of hidden
variable theories. Since the first experiment by Aspect {\it et al}.\cite
{Aspect}, many experiments have been performed to test Bell-type inequalities%
\cite{Experiments}. Most of the experiments showed that Bell-type
inequalities are violated, i.e., quantum mechanics is right. But because of
the low detection efficiency, nearly all of these experiments have to take
the fair sampling assumption except for a recent experiment by Rowe {\it et
al}.\cite{Rowe} who claimed that they had eliminated the ``detection''
loophole in their experiment. The contradiction between LHV and QM comes
from the fact that for LHV predetermined values must be assigned for
observables which are spacelike separated.

For another kind of hidden variables, which is called noncontextual hidden
variables (NCHV), an even more stringent demand is given out. That is, the
predetermined value for an observable in NCHV does not depend on the
experimental context, i.e., which comeasurable observables are measured
simultaneously, and spacelike separation is not needed between observables.
Without the requirement for spacelike separation between observables, NCHV
theories reveal a basic opinion in hidden variable theories more directly,
that is, the value we get from one measurement of an observable must be
predetermined, whether there are any other comeasurable observables being
measured simultaneously must not affect its value.

The Kochen-Specker (KS) theorem\cite{Kochen-Specker,Others,Cabello2} designs
a rather complex formulation to show that NCHV is not compatible with
quantum mechanics. But for a long time, there is not any experimental
disproof of NCHV theories using KS theorem (recently two experiments are
completed\cite{Michler} to test NCHV theories, but they use three particle
Greenberger-Horne-Zeilinger (GHZ) theorem and Bell-like inequality). The
reasons are pointed out by Cabello and Garcia-Alcaine (CG) in Ref. \cite{CG}%
: (i) The proof of the KS theorem refers to a single individual system but
involve noncompatible observables that cannot be measured in the same
individual system. (ii) The proof also refers to NCHV theories which share
some properties with quantum mechanics. So they are not entirely independent
of the formal structure of quantum mechanics.

In Ref. \cite{CG}, CG proposed an experimental scheme to test KS theorem
based on two spin-$\frac 12$ particles. They showed that the proof is
completely independent of the formal structure of quantum mechanics. In a
recent paper\cite{Simon}, Simon {\it et al}. present a rather simple
experimental scheme to prove that NCHV theories is not compatible with
quantum mechanics. Their scheme is feasible with single particles, using
both the path and spin degrees of freedom to form a two-qubit system.
Comparing with the experiments\cite{Experiments} that have been exhibited to
disprove the LHV theories (Except for a recent experiment by Pan {\it et al}.%
\cite{Pan}, which tests LHV in a nonstatistical way using three-photon
Greenberger-Horne-Zeilinger entanglement), the above two schemes are both
nonstatistical tests of the NCHV theories, they provide a very direct
all-or-nothing-type demonstration of the contradiction between NCHV theories
and QM. These all-or-nothing-type schemes make the real experiments easier
to discuss. And because the scheme by Simon et al.\cite{Simon} uses single
particles which makes it easier to be performed, we choose this scheme to
test the NCHV theories in experiment.

In the following paragraphs we will show our experimental setup and explain
it in detail.

In our experiment we use the polarization and path degrees of a single
photon to form a two-qubit system, that is different from the original
proposal in Ref. \cite{Simon} which uses spin-$\frac 12$ particles. This
difference does not affect our purpose at all, because photons and spin-$%
\frac 12$ particles are completely equivalent in this scheme.

Fig. 1 shows our experimental setup. The single photon source is provided by
one photon of the emitted photon pair produced through the spontaneous
parametric down-conversion process in a 1-mm thick type-I phase matching BBO
crystal, which is pumped by a $351.1$ nm laser beam ($100$ mW) produced by
an Ar$^{\text{+}}$ laser (Coherent, Sabre, model DBW25/7). The other photon
of the pair is detected as a trigger, and the coincidence rates are recorded
as the experimental data. This makes sure that the recorded data is provided
by single photons from the emitted photon pairs, not by other noises. A time
window of 5 ns is chosen to capture true coincidences and photons are
detected by single photon detectors ($D0\sim D8$)---silicon avalanche
photodiodes (EG\&G, SPCM-AQR), with efficiencies of $\sim 70\%$ at $702.2$
nm and dark counts of order $25$ s$^{\text{-1}}$, each is placed after a $%
4.6 $ nm interference filter (IF) and a $40\times $ lens.

To make Fig. 1 easier to discuss, we will now describe the scheme. In the
scheme four observables $Z_1,$ $X_1,$ $Z_2$ and $X_2$ are considered, where
the subscript $1,2$ denote the path and polarization qubit respectively.
Each of the observables has possible values $+1$ or $-1$. In the language of
quantum mechanics, we prepare a single photon in a two-qubit state $\left|
\Psi \right\rangle $ using its path and polarization degree 
\begin{equation}
\left| \Psi \right\rangle =\frac 1{\sqrt{2}}(\left| u\right\rangle \left|
z+\right\rangle +\left| d\right\rangle \left| z-\right\rangle ),  \eqnum{1}
\end{equation}
where $\left| u\right\rangle $ and $\left| d\right\rangle $ denote the
``up'' and ``down'' paths of the photon after a beamsplitter, $\left|
z+\right\rangle $ and $\left| z-\right\rangle $ denote the ``vertical'' and
``horizontal'' polarization states of the photon. The four observables $Z_1,$
$X_1,$ $Z_2$ and $X_2$ are represented by\cite{Simon}

\begin{eqnarray}
Z_1 &=&\left| u\right\rangle \left\langle u\right| -\left| d\right\rangle
\left\langle d\right| ,  \nonumber \\
X_1 &=&\left| u^{\prime }\right\rangle \left\langle u^{\prime }\right|
-\left| d^{\prime }\right\rangle \left\langle d^{\prime }\right| ,  \eqnum{2}
\\
Z_2 &=&\left| z+\right\rangle \left\langle z+\right| -\left| z-\right\rangle
\left\langle z-\right| ,  \nonumber \\
X_2 &=&\left| x+\right\rangle \left\langle x+\right| -\left| x-\right\rangle
\left\langle x-\right| ,  \nonumber
\end{eqnarray}
where $\left| u^{\prime }\right\rangle =\frac 1{\sqrt{2}}(\left|
u\right\rangle +\left| d\right\rangle ),$ $\left| d^{\prime }\right\rangle =%
\frac 1{\sqrt{2}}(\left| u\right\rangle -\left| d\right\rangle ),$ $\left|
x+\right\rangle =\frac 1{\sqrt{2}}(\left| z+\right\rangle +\left|
z-\right\rangle ),$ $\left| x-\right\rangle =\frac 1{\sqrt{2}}(\left|
z+\right\rangle -\left| z-\right\rangle )$.

We can learn from $\left| \Psi \right\rangle $ that the measurement of $%
Z_1Z_2$ and $X_1X_2$ will both get the results of $+1$. And when we perform
the joint measurement of $Z_1X_2$ and $X_1Z_2$, QM predicts that the results
of them will definitely be opposite\cite{Simon}.

On the other hand, in the language of NCHV theories, the scheme can be
described as below: First we prepare many systems (the single photons) in a
certain way and show that for four observables $Z_1,$ $Z_2,$ $X_1$, $X_2$
(defined by certain experimental operations) these systems all have the
property that the results of measurements of $Z_1Z_2$ and $X_1X_2$ both
equal to $+1$. Then for systems prepared in the same way, we perform the
joint measurement of $Z_1X_2$ and $X_1Z_2$. NCHV theories predict that the
results of $Z_1X_2$ and $X_1Z_2$ will surely be equal\cite{Simon} and this
leads to the contradiction with QM. So our task is to prepare such systems
and perform a joint measurement of $Z_1X_2$ and $X_1Z_2$ to determine
whether QM or NCHV theories give the right answer. In the following we will
show that the setup in Fig. 1 can be used for this task.

In Fig. 1, one photon of the pair is detected by $D0$ as a trigger. The
other photon is prepared in the required state (described as $\left| \Psi
\right\rangle $ in QM) by a properly rotated half-wave plate (HWP0), and a
polarizing beamsplitter (PBS0)---each PBS in Fig. 1 is set to reflect
vertical polarization photons. For the symmetry of the setup in Fig. 1, in
the following we will first discuss the interferometer formed by PBS0 and
BS1 in detail and then briefly discuss the other interferometer formed by
PBS0 and BS2.

To make things clear in NCHV theories, now we shall give the definitions of $%
Z_1,$ $Z_2,$ $X_1$ and $X_2$ in an operational way. Corresponding to the
definitions in Equation $(2)$, we can easily get that: (i). $Z_1$ means in
which path after PBS0 we find the photon, ``up ($+1$)'' or ``down ($-1$)''.
(ii). $Z_2$ is the measurement of polarization of the photon, ``vertical ($%
+1 $)'' or ``horizontal ($-1$)''. (iii). $X_1$ means that when the ``up''
and ``down'' path length between PBS0 and BS1 equals, we find the photon in
``up'' or ``down'' path after BS1 (because the interference on a BS performs
a Hardmard transformation of the path qubit). (iv). $X_2$ means that the
photon is polarized at $+45^{\circ }$ ($+1$) or $-45^{\circ }$ ($-1$) away
from the horizontal direction.

From the above definitions, we can easily see that the property $Z_1Z_2=+1$
of the prepared photons is decided by the performance of HWP0 and PBS0.
Because the rather good quality of our PBS'es (extinction ratio of the order 
$10^{-5}$) and HWPs ($\Delta \theta =0.2^{\circ }$), we can regard that the
prepared photons just have the property $Z_1Z_2=+1$.

To decide whether $X_1X_2=+1$ or $-1$ for our prepared single photons, an
equal-arm interferometer formed by PBS0 and BS1 is used to measure $X_1$;
HWP3 (set at $+22.5^{\circ }$) followed by PBS3 both with HWP4 (set at $%
+22.5^{\circ }$) followed by PBS4 are used to measure $X_2$. While working
together, they can measure $X_1X_2$. When measuring $X_1X_2$, HWP1 and HWP2
are set at $0^{\circ }$ to just let the photons pass without changing its
polarization state. We name the above setup as Setup1 in the following.

The setup of joint measurement of $Z_1X_2$ and $X_1Z_2$ is very similar with
Setup1, the only difference is that HWP1 and HWP2 are respectively set at $%
+22.5^{\circ }$ and $-67.5^{\circ }$ to perform the measurement of $X_2$
(with PBS1 and PBS2). In the original setup\cite{Simon} two more HWPs should
be placed, one between PBS1 and BS1, the other between PBS2 and BS1 to
rotate the polarization of the photon to $+45^{\circ }$ and $-45^{\circ }$
respectively for the measurement of $X_2$ (In fact, we also need two more
HWPs between PBS1, PBS2 and BS2 for the same reason. But we will not discuss
it because of their similarity). On the other hand, HWP3 and HWP4 should be
set at angle $0^{\circ }$ to measure $Z_2$. But in the concept of a $%
+45^{\circ }$ rotation of our polarization measurement basis, the setup that
HWP3 and HWP4 at $+22.5^{\circ }$ (without two more HWPs) will not result in
any change in the recorded data compared with the original setup. So, the
setup of HWP1 and HWP2 respectively at $+22.5^{\circ }$ and $-67.5^{\circ }$
will complete the last step in our task, that is, to decide whether $Z_1X_2$
equals to $X_1Z_2$ or not. We name this setup as Setup2. We note here that
it is important to make sure that in Setup2 the two interferometers formed
by PBS0 and BS1 (BS2) are both ``equal-arm''. The measurement of $X_1X_2$
using Setup1 can tell us whether the interferometer formed by PBS0 and BS1
is ``equal-arm'' in Setup1, so we only need to smoothly change the setup
from Setup1 to Setup2 to make sure the ``equal-arm'' property of the
interferometer of PBS0 and BS1. For the other interferometer (PBS0 and BS2),
we can use a similar method, i.e., we set HWP1 and HWP2 both at $+45^{\circ
} $ to make all the photons change their way to BS2. Then we can use the
interferometer (BS2) to measure $X_1X_2$ (HWP5 and HWP6 both set at $%
+22.5^{\circ }$) and make sure its ``equal-arm'' property in Setup2. We call
this setup as Setup1'.

It can be verified\cite{Simon} that if QM is right, only $D1,$ $D3,$ $D5,$ $%
D7$ will detect photons in Setup2 when only $D2,$ $D4$ $(D6,$ $D8)$ have
detected photons in Setup1 (Setup1'). While for NCHV theories only $D2,$ $%
D4, $ $D6,$ $D8$ will detect photons in Setup2. Thus this is an
all-or-nothing-type experiment.

Because of the limitation of our devices, we didn't record all the
coincidence rates of $D0$ and $D1\sim D8$ at the same time. Instead, each
time we only record one of them. But the experiment process of the eight
cases are almost the same.

Now we will describe the experiment process.

For the first step we use Setup1 (Setup1'). The interferometer's arm-length
is tuned so that it reaches its minimal value at $D1,$ $D3,$ $D5,$ $D7$
(while reaches its maximal value at $D2,$ $D4,$ $D6,$ $D8$ ). This result
shows that the prepared photons have the property of $X_1X_2=+1,$ which can
be easily deduced from the operational definitions for the four observables
above (For example, the minimal value at $D1$ can be regarded as no photon
detected by $D1$ in spite of the imperfections in experiment. So, $X_1=-1$
and $X_2=-1$). In the second step, we smoothly change Setup1 (or Setup1' for
the case of $D5,$ $D6,$ $D7,$ $D8$) to Setup2, just by rotating HWP1 and
HWP2 to $+22.5^{\circ }$ and $-67.5^{\circ }$ respectively with a specially
designed mechanical device, and record the correspondent coincidence rates.
If now the interferometer reached its maximal (minimal) value in $D1,$ $D3,$ 
$D5,$ $D7$ ($D2,$ $D4,$ $D6,$ $D8$), we know that $Z_1X_2=-1$ $(+1)$ and $%
X_1Z_2=+1$ $(-1)$\cite{Step2}, that means QM is right.

The recorded coincidence rates between $D0$ and $D1\sim D8$ are shown in
Fig. 2, from $D1$ to $D8$. The vertical axis is the recorded coincidence
rates and the horizontal axis is the time axis of performing the
experiments. We can see that each figure in Fig. 2 has three stages along
the time axis. In the first stage, the setup is in Setup1 (or Setup1 for $%
D5, $ $D6,$ $D7,$ $D8$) and the coincidence rates are stable at the
interferometer's minimal value for $D1,$ $D3,$ $D5,$ $D7$ maximal value for $%
D2,$ $D4,$ $D6,$ $D8$. That shows $X_1X_2=+1$ for the prepared photons. In
the second stage, the setup has been changed to Setup2 (The time of changing
setup is only about $2s$), and the recorded coincidence rates are stable at
the interferometer's maximal value\cite{Maximum} for $D1,$ $D3,$ $D5,$ $D7$
or minimal value for $D2,$ $D4,$ $D6,$ $D8$, that shows $Z_1X_2$ is opposite
to $X_1Z_2$. In the third stage, the setup is again in Setup1 (Setup1') and
the recorded rates recover to the same level as those in the first stage,
that means our system is stable and controllable during the experiment (In
fact, the stable time of the interferometer is about $5$ minutes).

Fig. 3 is the analyzed experiment results. It shows to what extent our
results violate the prediction of NCHV theories.

As Simon {\it et al}. pointed out in Ref. \cite{Simon}, the appearance of
the paradox in this experiment is related to the superposition principle, so
choosing the right angle of HWP1 and HWP2 is very important. If we set HWP1
and HWP2 both at $+22.5^{\circ }$ or $-67.5^{\circ }$ in Setup2, theoretical
analysis and further experiment show that the results would not lead to
contradiction between QM and NCHV theories.

An important problem which may lead to argument on this experiment is that
whether the finite precision measurement would nullify the demonstration in
this paper. This kind of problem in Kochen-Specker-type experiments was
first argued by Meyer\cite{Meyer}, who claimed that the Kochen-Specker
theorem was ``nullified'' in real experiments because of the unavoidably
finite measurement precision. Kent\cite{Kent} generalized his work and came
to a similar conclusion.

In response, Simon {\it et al}.\cite{Simon2} demonstrated in an entirely
operational way that even when the finite measurement precision is taken
into account, NCHV theories still can be excluded from Kochen-Specker
theorem. In a recent paper\cite{Cabello}, Cabello also proved that only
finite measurement precision is needed to disprove NCHV theories in real
Kochen-Specker experiments. Because in our experiment only four observables $%
Z_1$, $X_1$, $Z_2$, $X_2$ are measured, following the analysis in Ref. \cite
{Simon2}, the Kochen-Specker set is 
\begin{equation}
\{\{Z_1,Z_2\},\{X_1,X_2\},\{Z_1X_2,X_1Z_2\}\}.  \eqnum{3}
\end{equation}
The contradiction comes from the fact that the set is constructed in such a
way that there is no way to assign predetermined values to $Z_1$, $X_1$, $%
Z_2 $, $X_2$ to make the measurement values of the pairs in set $(3)$ all
satisfy the predictions of QM. The number of measurements in the set is $N=3$%
. So the error fraction $\epsilon \leq \frac 13$ can satisfy the requirement
to disprove NCHV theories in this experiment, that is, the photons that
violate the prediction of QM must be less than $\frac 13$ of all photons.
And Fig. 3 has shown that $\epsilon $ is about $0.19$ in our experiment. In
this way, we can still come to our conclusion that this experiment disproves
the NCHV theories in a nonstatistical way.

In conclusion, we have performed an all-or-nothing test of NCHV theories
using a simple Kochen-Specker theorem model and the results proves that
quantum mechanics is right.

This work was funded by National Fundamental Research Program
(2001CB309300), National Natural Science Foundation of China, the Innovation
funds from Chinese Academy of Sciences, and also by the outstanding Ph. D
thesis award and the CAS's talented scientist award entitled to Luming Duan.
This work was also supported by the CAS's (ChineseAcademy of Sciences)
talented young scientist award entitled to Jian-Wei Pan.

{\bf Figure Captions:}

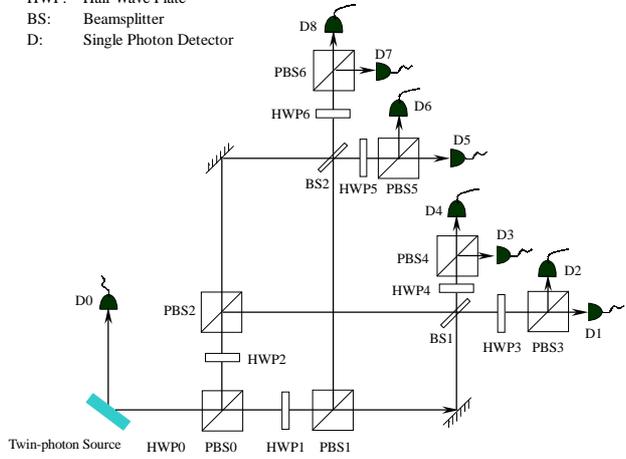
Figure 1: Experimental setup of Setup1 (Setup1') and Setup2. For Setup1
(Setup1'), HWP1 and HWP2 are both set at $0^{\circ }$ ($+45^{\circ }$). For
Setup2, HWP1 is set at $+22.5^{\circ }$, and HWP2 is set at $-67.5^{\circ }$.

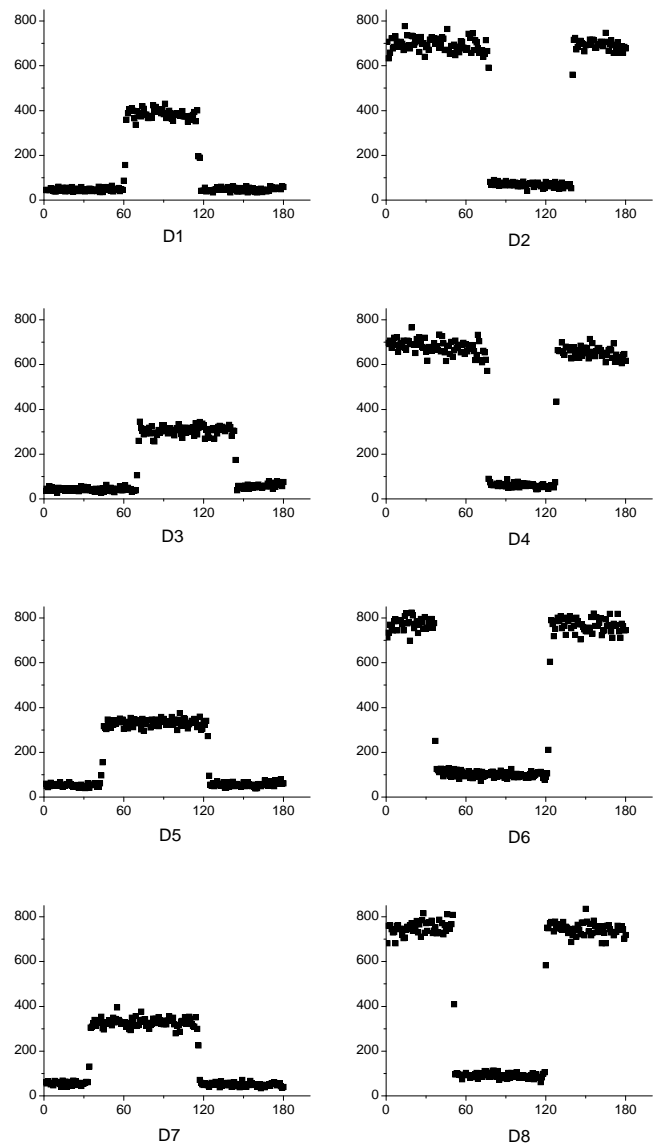
Figure 2: Recorded coincidence rates of $D0$ and $D1\sim D8$ in the
experiment process. The vertical axis is the coincidence rates (s$^{\text{-1}%
}$) and the horizontal axis is the time axis (s) of performing the
experiment.

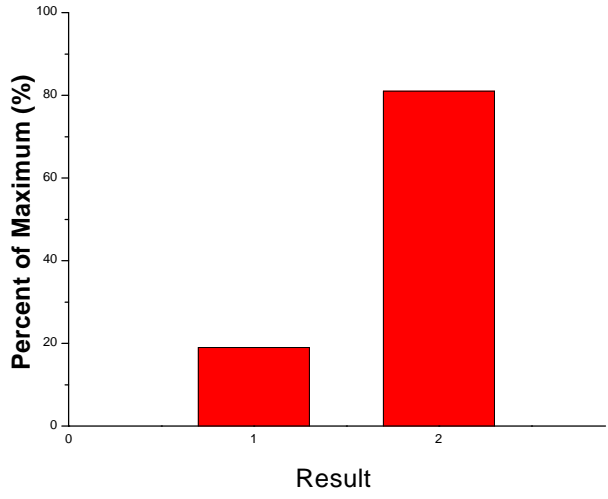
Figure 3: Analyzed experiment results. Result 1 is the fraction of total
coincidence rates that agree with QM, that is the probability of finding the
photon in $D1,$ $D3,$ $D5,$ or $D7$ in Setup2. Result 2 is the fraction that
agree with NCHV, that is the probability of finding the photon in $D2,$ $D4,$
$D6,$ or $D8$ in Setup2. The difference between result 1 and result 2 shows
the violation against NCHV theories in our experiment.


\begin{references}
\bibitem{Step2}  In detail, when $D1$ gets maximum value, we can conclude
the values of the four observables: $Z_1=+1$ $(\left| u\right\rangle ),$ $%
X_2=-1$ $(-45^{\circ })$ or $Z_1=-1$ $(\left| d\right\rangle ),$ $X_2=+1$ $%
(+45^{\circ }),$ that is, $Z_1X_2=-1$; and $X_1=-1$ $(\left| d^{\prime
}\right\rangle )$, $Z_2=-1$ $($Horizontal$),$ which means $X_1Z_2=+1$.
\bibitem{Maximum}  In Setup2 the maximal value in $D1,$ $D3$($D5,$ $D7$) of
the interferometer is only half of that in Setup1(Setup1') because half of
the photons are detected by $D5\sim D8$($D1\sim D4$). The maximal value of $%
D1,$ $D3$($D5,$ $D7$) in Setup1(Setup1') is about $750$ $s^{-1}$, so the
maximal value in Setup2 is about $375$ $s^{-1}$.
\end{references}
\end{document}